\documentclass[12pt,aps,showpacs,eqsecnum,nofootinbib,floatfix]{revtex4}
\usepackage{bbding}
\usepackage{pifont}
\usepackage{subfigure}
\usepackage{epsfig}
\usepackage{amsmath}
\usepackage{amsfonts}
\usepackage{amssymb}
\usepackage{color}
\usepackage{graphicx}
\usepackage{mathrsfs}
\usepackage{multirow}

\def\be{\begin{equation}}
\def\ee{\end{equation}}
\def\ba{\begin{eqnarray}}
\def\ea{\end{eqnarray}}

\newcommand{\omits}[1]{}

\begin{document}
%\begin{CJK*}{GBK}{song}

\title{Two-phase equilibrium properties in charged topological dilaton AdS black holes}

\author{Hui-Hua Zhao, Li-Chun Zhang, Ren Zhao\footnote{corresponding author: Email:kietemap@126.com(Ren Zhao)}}

\medskip

\medskip

\affiliation{Institute of Theoretical Physics, Shanxi Datong University,
Datong 037009, China}
\affiliation{Department of Physics, Shanxi Datong University,
Datong 037009, China}

\begin{abstract}

In this paper we discuss phase transition of the charged topological dilaton AdS black holes by Maxwell equal area law. The two phases involved in the phase transition could be coexist and we depict the coexistence region in $P-v$ diagrams. The two-phase equilibrium curves in $P-T$ diagrams are plotted, the Clapeyron equation for the black hole is derived, and the latent heat of isothermal phase transition is investigated. We also analyze of the parameters of the black hole that have an effect on the two phases coexistence.  The results show that the black hole may go through a small-large phase transition similar to those of usual non-gravity thermodynamic systems.


\end{abstract}

\maketitle

\bigskip

\section{Introduction}

In recent years, the cosmological constant in $n$-dimensional AdS and dS
spacetime has been regarded as pressure of black hole
thermodynamic system with
\begin{equation}
\label{eq1} P=-\frac{\Lambda }{8\pi }\quad,
\end{equation}
and the corresponding conjugate quantity, thermodynamic
volume\cite{Kastor:2009wy,Cvetic,Guna,Dola}
\begin{equation}
\label{eq2}
V=\left( {\frac{\partial M}{\partial P}} \right)_{S,Q_i ,J_k }.
\end{equation}
The $(P\sim V)$ critical behaviors in AdS and dS black holes have
been extensively studied\cite{David,Frass,David1,Altami,Altami1,Altami2,Banerjee,
Ghaff,Zhao,Zhao1,Zhao2,Ma,Zhang,Ma1,Ma2,Heidi,Arci,Mo,Mo1,Mo2,Mo3,Mo4,Mo5,Lala,Man,
Wei,Suresh,Mans,Thar,Niu,Ma3,Cai,Zou,Zou1,Li,Wei1,Poshteh,Xu,Liu,Xu1}. Using Ehrenfest
scheme, Ref.\cite{Mo,Mo1,Mo2,Mo3,Mo4,Mo5,Lala,Man} studied the critical phenomena in
a series of black holes in AdS spacetime, and proved the phase transition at
critical point is the second order one, which has also been confirmed in
Ref.\cite{Wei,Suresh,Mans,Thar,Niu} by studying thermodynamics and state space geometry of black
holes. And a completely simulated gas-liquid
system has been put forward\cite{Guna,David,Altami,Liu}. Recently phase transition below critical temperature and
phase structure of some black holes have received much
attention\cite{Wei2,Zhang2,ZhaoHui,Belhaj,Hao}.

Although some encouraging results about black hole thermodynamic
properties in AdS and dS spacetimes have been achieved and the problems
about phase transition of black holes have been extensively discussed,
an unified recognition about the phase transition of black hole has not
 been put forward.It is significant to further explore phase
equilibrium and phase structure in black holes, which can
help to recognize the evolution of  black hole. We also expect to
provide some relevant information for exploring quantum gravity
properties by studying the phase transition of charged topological
dilaton AdS black holes.

A scalar field called dilaton appears in the low energy limit of
string theory. The presence of the dilaton field has important
consequences on the causal structure and the thermodynamic
properties of black holes. Much interest has been focused on
studies of the dilaton black holes in recent
years\cite{Gold,Chen,Gout,Lizuka,LiWJ,LiWJ1,Gubser,Shey,Gold1,Ong,Shey1}.
The isotherms in $P\sim v$ diagrams of charged topological dilaton
AdS black hole in Ref.\cite{Zhao} show there exists thermodynamic
unstable region with $\partial P/\partial v>0$ when temperature is
below critical temperature and the negative pressure emerges when
temperature is below a certain value. This situation also exists in
van der Waals-Maxwell gas-liquid system, which has been resolved by
Maxwell equal area law. In this paper, using the Maxwell equal area law,
we establish an  phase transition process in charged topological dilaton
AdS black holes, where the issues about unstable states and negative pressure
are resolved. By studying the phase transition process, we acquire the two-phase
equilibrium properties including the $P-T$ phase diagram, Clapeyron equation and latent heat of phase change.
The results show the simulated phase transition is the first order phase transition but phase transition
at critical point belongs to the continuous one though the the parameters
of the charged topological dilaton black hole that have some effects on
the two phases coexistence.

The paper is arranged as follow. The charged topological dilaton AdS
black hole as a thermodynamic system is briefly introduced in
section 2. In section 3, by Maxwell equal area law the phase
transition processes at certain temperatures are obtained and the
boundary of two phase equilibrium region are depicted in $P-v$
diagram for a charged topological dilaton AdS black hole. Then
some parameters of the black hole are analyzed to find the relevance
with the two-phase equilibrium. In section 4, the $P-T$ phase diagrams are plotted
and the Clapeyron equation and latent heat of the phase change are derived.
We make some discussions and conclusions in section 5.
we use the units $G_d =\hbar =k_B=c=1$ in this paper)

\section{Charged Dilaton Black Holes in Anti-de Sitter Space}

The Einstein-Maxwell-Dilaton action in $(n+1)$-dimensional $(n\ge
3)$spacetime is\cite{Ong,Shey1}
\begin{equation}
\label{eq3} S=\frac{1}{16\pi }\int {d^{n+1}} x\sqrt {-g} \left(
{R-\frac{4}{n-1}(\nabla \Phi )^2-V(\Phi )-e^{-4\alpha \Phi
/(n-1)}F_{\mu \nu } F^{\mu \nu }} \right),
\end{equation}
where the dilaton potential is expressed in terms of the dilaton field and
its coupling to the cosmological constant:
\begin{equation}
\label{eq4} \nabla ^2\Phi =\frac{n-1}{8}\frac{\partial U}{\partial
\Phi }-\frac{\alpha }{2}e^{-4\alpha \Phi /(n-1)}F_{\lambda \eta }
F^{\lambda \eta },
\end{equation}
\begin{equation}
\label{eq5}
\nabla _\mu \left( {e^{-4\alpha \Phi /(n-1)}F^{\mu \nu }} \right)=0,
\end{equation}
where $R$ is the Ricci scalar curvature, $\Phi $ is the dilaton field and
$V(\Phi )$ is a potential for $\Phi $, $\alpha $ is a constant determining
the strength of coupling of the scalar and electromagnetic field, $F_{\mu
\nu } =\partial _\mu A_\nu -\partial _\nu A_\mu $ is the electromagnetic
field tensor and $A_\mu $ is the electromagnetic potential. The topological
black hole solutions take the form\cite{Ong,Shey1}
\begin{equation}
\label{eq6}
ds^2=-f(r)dt^2+\frac{dr^2}{f(r)}+r^2R^2(r)d\Omega _{k,n-1}^2 ,
\end{equation}
where
\[
f(r)=-\frac{k(n-2)(\alpha ^2+1)^2b^{-2\gamma }r^{2\gamma }}{(\alpha
^2-1)(\alpha ^2+n-2)}-\frac{m}{r^{(n-1)(1-\gamma )-1}}+\frac{2q^2(\alpha
^2+1)^2b^{-2(n-2)\gamma }}{(n-1)(\alpha ^2+n-2)}r^{2(n-2)(\gamma -1)}
\]
\begin{equation}
\label{eq7}
-\frac{n(\alpha ^2+1)^2b^{2\gamma }}{l^2(\alpha ^2-n)}r^{2(1-\gamma )},
\end{equation}
\begin{equation}
\label{eq8}
R(r)=e^{2\alpha \Phi /(n-1)},
\quad
\Phi (r)=\frac{(n-1)\alpha }{2(1+\alpha ^2)}\ln \left( {\frac{b}{r}}
\right),
\end{equation}
with $\gamma =\alpha ^2/(\alpha ^2+1)$ and $b$ is an arbitrary constant. The
cosmological constant is related to spacetime dimension $n$ by
\begin{equation}
\label{eq9}
\Lambda =-\frac{n(n-1)}{2l^2},
\end{equation}
where $l$ denotes the AdS length scale. In (\ref{eq7}), $m$ appears
as an integration constant and is related to the ADM
(Arnowitt-Deser-Misnsr) mass of the black hole. According to the
definition of mass due to Abbott and Deser, the ADM mass of the solution
(\ref{eq7}) is
\begin{equation}
\label{eq10}
M=\frac{b^{(n-1)\gamma }(n-1)\omega _{n-1} }{16\pi (\alpha ^2+1)}m
\end{equation}
The electric charge is
\begin{equation}
\label{eq11}
Q=\frac{q\omega _{n-1} }{4\pi },
\end{equation}
where $\omega _{n-1} $ represents the volume of constant curvature
hypersurface described by $d\Omega _{k,n-1}^2 $

The thermodynamic quantities satisfy the first law of thermodynamics
\begin{equation}
\label{eq12}
dM=TdS+UdQ+VdP
\end{equation}
The Hawking temperature and entropy of the topological black hole
\begin{equation}
\label{eq13}
T=-\frac{(\alpha ^2+1)}{2\pi (n-1)}\left( {\frac{k(n-2)(n-1)b^{-2\gamma
}}{2(\alpha ^2-1)}r_+^{2\gamma -1} +\Lambda b^{2\gamma }r_+^{1-2\gamma }
+q^2b^{-2(n-2)\gamma }r_+^{(2n-3)(\gamma -1)-\gamma } } \right)
\end{equation}
\begin{equation}
\label{eq14}
S=\frac{b^{(n-1)\gamma }\omega _{n-1} r_+^{(n-1)(1-\gamma )} }{4}
\end{equation}
where $r_+ $ represents the position of black hole horizon and meets
$f(r_+ )=0$. The electric potential
\begin{equation}
\label{eq15}
U
=\frac{qb^{(3-n)\gamma }}{r_+^\lambda \lambda },
\end{equation}
and the pressure and volume are respectively
\begin{equation}
\label{eq16}
P=\frac{n(n-1)}{16\pi l^2},
\quad
V=-\frac{(\alpha ^2+1)b^{\gamma (n+1)}\omega _{n-1} }{(\alpha
^2-n)}r_+^{n-\gamma (n+1)}
\end{equation}
where $\lambda =(n-3)(1-\gamma )+1$.

Using the Eqs. (\ref{eq9}), (\ref{eq13}) and (\ref{eq16}) for a
fixed charge Q, one may obtain the equation of state $P(v,T)$,
\[
P=\frac{T}{v}
+\frac{k(n-2)(\alpha ^2+1)^2}{\pi (n-1)(\alpha ^2-1)v^2}
+\frac{Q^2b^{2(1-n)\gamma }2\pi }{\omega _{n-1}^2 }\left(
{\frac{v(n-1)}{4(\alpha ^2+1)b^{2\gamma }}}
\right)^{\textstyle{{2(n-1)(\gamma -1)} \over {1-2\gamma }}}
\]
\begin{equation}
\label{eq17}
=\frac{T}{v}
-\frac{A}{v^2}
+\frac{B}{v^{\mbox{2}(n-\mbox{1})(\mbox{1}-\gamma )/(1-\mbox{2}\gamma )}},
\end{equation}
where specific volume [12]
\begin{equation}
\label{eq18}
v=\frac{4(\alpha ^2+1)b^{2\gamma }}{(n-1)}r_+^{1-2\gamma } ,
\end{equation}
and
\[
d=\frac{2(n-1)(1-\gamma )}{1-2\gamma },
\quad
A=\frac{k(n-2)(\alpha ^2+1)^2}{\pi (n-1)(1-\alpha ^2)}
\]
\begin{equation}
\label{eq19}
B=\frac{Q^2b^{2(1-n)\gamma }2\pi }{\omega _{n-1}^2 }\left( {\frac{4(\alpha
^2+1)b^{2\gamma }}{(n-1)}} \right)^{2(n-1)(1-\gamma )/(1-2\gamma )}.
\end{equation}
In Fig.1 we plot the isotherms in $P-v$ diagrams in terms of state equation Eq.
(\ref{eq17}) at different dimension $n$, charge $Q$, and parameters
$b$ and $\alpha $. One can see from Fig.1 that there are
thermodynamic unstable segments with $\partial P/\partial v>0$ on
the isotherms as temperature $T<T_c $, where $T_c $ is critical
temperature. And the negative pressure emerges when temperature is
below certain value $\tilde {T}$. $\tilde {T}$ and the corresponding
specific volume $\tilde {v}$ can be derived.
\begin{equation}
\label{eq20}
\tilde {v}^{d-2}=\frac{B}{A}(d-1),
\quad
\tilde {T}=\frac{A\left( {d-2} \right)}{\tilde {v}\left( {d-1} \right)}.
\end{equation}

\begin{figure}[!htbp]
\center{\subfigure[~$n=3$,$b=0.5$,$Q=2$,$\alpha=0.01$] {
\includegraphics[angle=0,width=5cm,keepaspectratio]{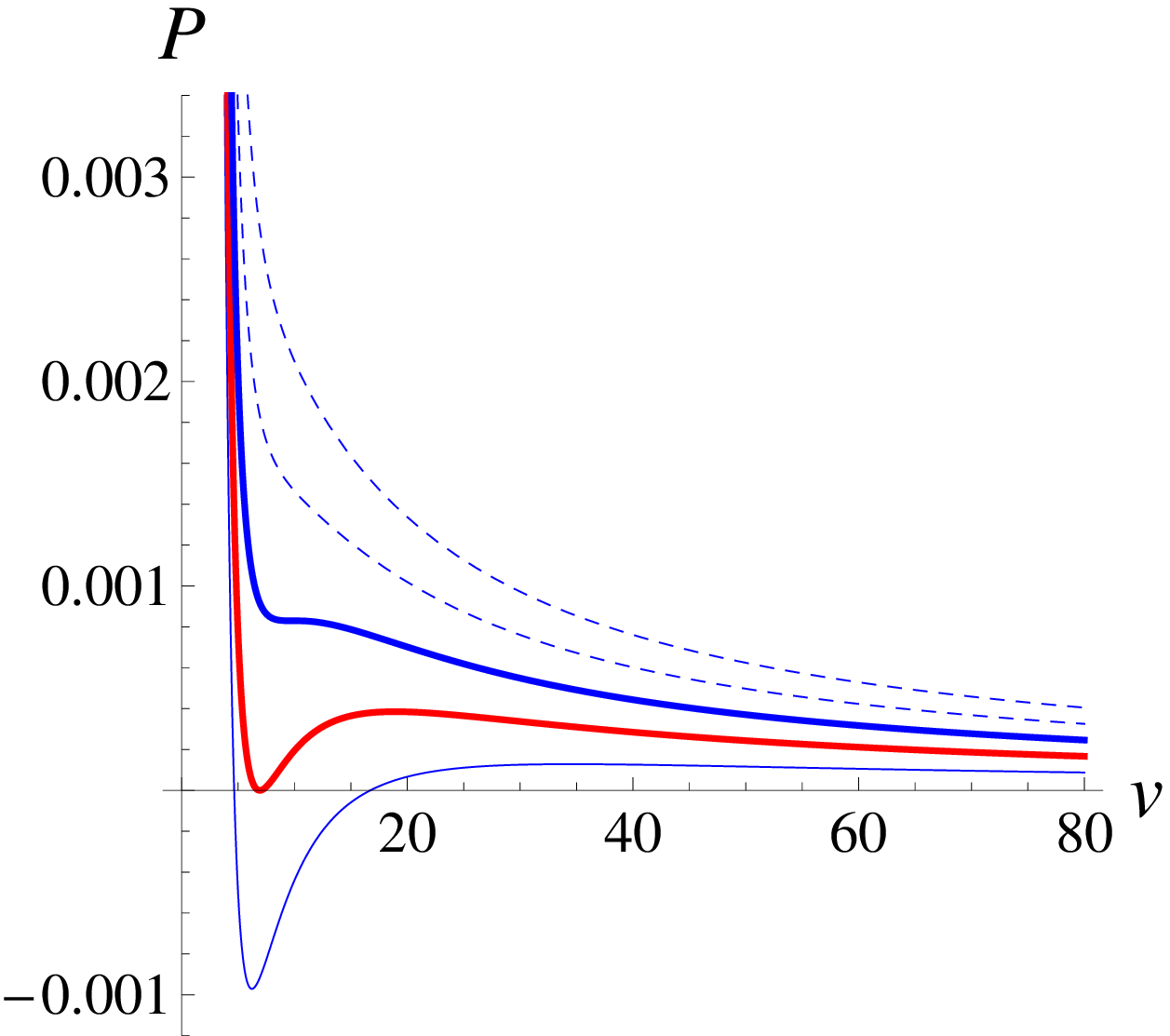}}
\subfigure[~$n=5$,$b=1$,$Q=10$,$\alpha=0.1$] {
\includegraphics[angle=0,width=5cm,keepaspectratio]{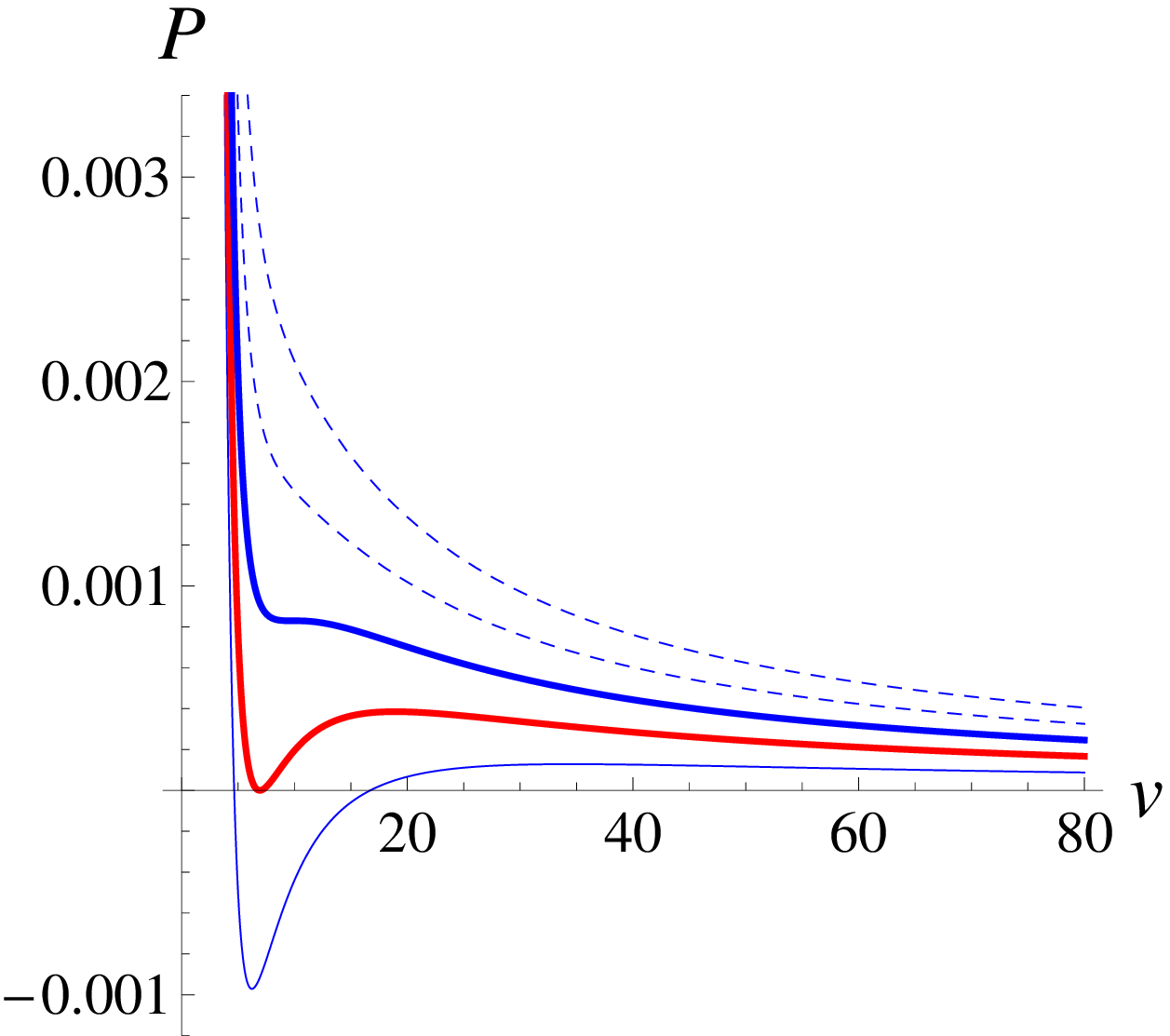}}
\subfigure[~$n=7$,$b=10$,$Q=30$,$\alpha=0.5$] {
\includegraphics[angle=0,width=5cm,keepaspectratio]{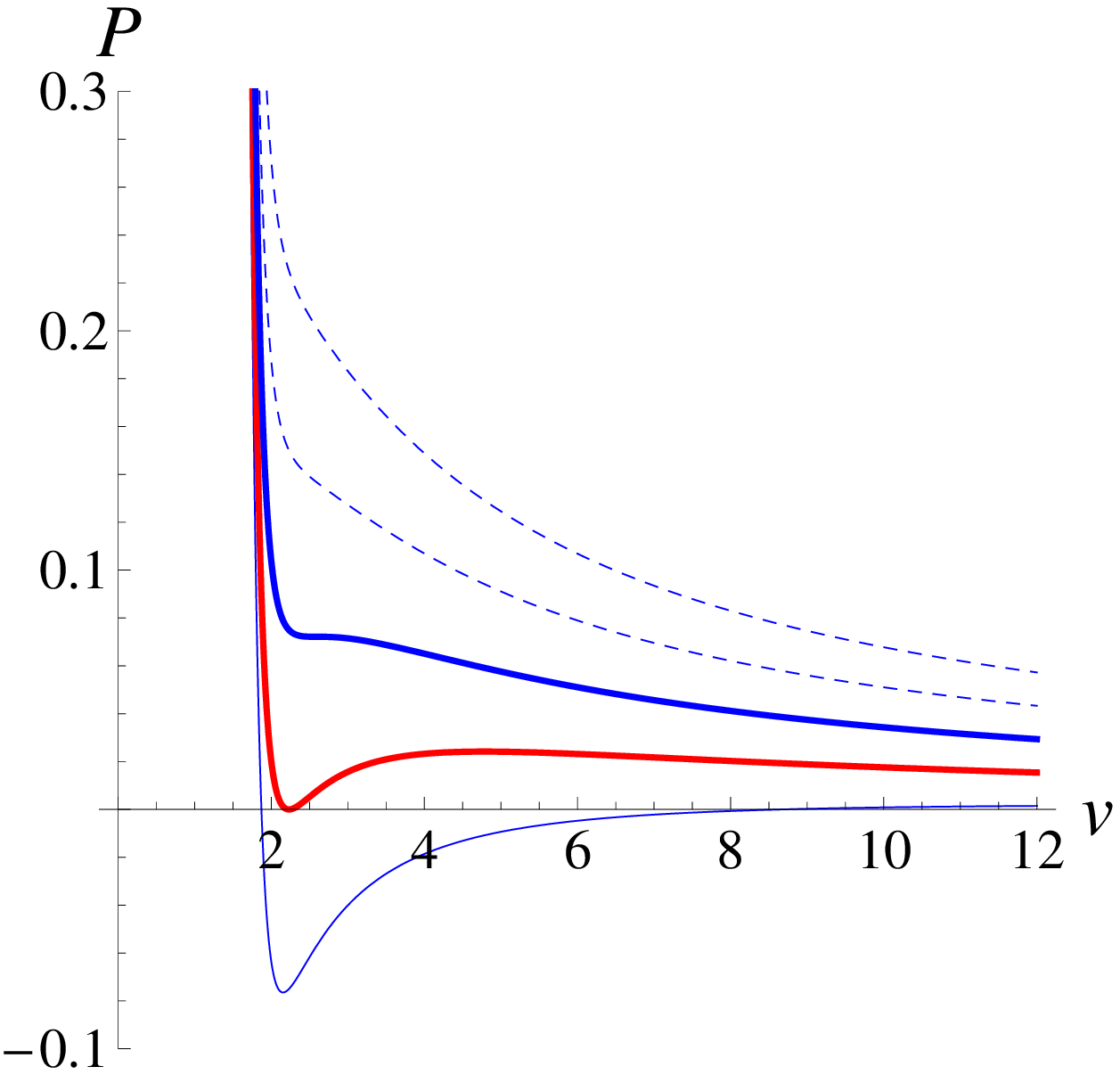}}
\caption[]{\it Isotherms in $P-v$ diagrams of charged topological dilaton black holes
in $n$ dimensional AdS spacetime}} \label{Pv1}
\end{figure}

\section{two-Phase equilibrium and Maxwell equal area law}

The state equation of the charged topological black hole is exhibited by the
isotherms in Fig.1, in which the thermodynamic unstable states
with $\partial P/\partial v>0$ will lead to the system
expansion or contraction automatically and the negative pressure
situation have no physical meaning. The cases occur also in van der
Waals equation but they have been resolved by Maxwell equal area law.

We extend the Maxwell equal area law to $n+1$-dimensional charged
topological dilaton AdS black hole to establish an phase transition process of the black
hole thermodynamic system. On the isotherm with temperature $T_0 $ in $P-v$
diagram, the two points $\left( {P_0 ,\;v_1 } \right)$ and $\left( {P_0
,\;v_2 } \right)$ meet the Maxwell equal area law,
\begin{equation}
\label{eq21}
P_0 (v_2 -v_1 )=\int\limits_{v_1 }^{v_2 } {Pdv} ,
\end{equation}
which results in
\begin{equation}
\label{eq22}
P_0 (v_2 -v_1 )=T_0 \ln \left( {\frac{v_2 }{v_1 }} \right)-A\left(
{\frac{1}{v_1 }-\frac{1}{v_2 }} \right)
+\frac{B}{d-1}\left( {\frac{1}{v_1^{d-1} }-\frac{1}{v_2^{d-1} }} \right),
\end{equation}
where the two points $\left( {P_0 ,\;v_1 } \right)$ and $\left( {P_0 ,\;v_2
} \right)$ are seen as endpoints of isothermal phase transition. Considering
\begin{equation}
\label{eq23}
P_0 =\frac{T_0 }{v_1 }-\frac{A}{v_1^2 }+\frac{B}{v_1^d },
\quad
P_0 =\frac{T_0 }{v_2 }-\frac{A}{v_2^2 }+\frac{B}{v_2^d },
\end{equation}
and setting $x=v_1 /v_2 $, we can get
\begin{equation}
\label{eq24}
T_0 v_2^{d-1} x^{d-1}=Av_2^{d-2} x^{d-2}(1+x)-B\frac{1-x^d}{1-x},
\end{equation}
\begin{equation}
\label{eq25}
P_0 x^{d-1}v_2^d =Ax^{d-2}v_2^{d-2} -B\frac{1-x^{d-1}}{1-x}\quad ,
\end{equation}
\begin{equation}
\label{eq26}
v_2^{d-2} =\frac{B}{A}\frac{d(1-x^{d-1})(1-x)+(d-1)(1-x^d)\ln
x}{x^{d-2}(d-1)(1-x)\left( {2(1-x)+(1+x)\ln x} \right)}=f(x).
\end{equation}
Substituting (\ref{eq26}) into (\ref{eq24}) and setting $T_0 =\chi T_c $ ($0<\chi <1)$, we
obtain
\begin{equation}
\label{eq27}
\chi T_c x^{d-1}f^{(d-1)/(d-2)}(x)=Af(x)x^{d-2}(1+x)-B\frac{1-x^d}{1-x}.
\end{equation}
When $x\to 1$, the corresponding state is critical point state. From (\ref{eq26})
\begin{equation}
\label{eq28}
v_2^{d-2} =v_1^{d-2} =v_c^{d-2} =f\left( 1 \right)=\frac{d(d-1)B}{2A}
\end{equation}
Substituting (\ref{eq28}) into (\ref{eq24}) and (\ref{eq25}), the
critical temperature and critical pressure are
\begin{equation}
\label{eq29}
T_c =\frac{2A(d-2)}{(d-1)}\left( {\frac{2A}{d(d-1)B}} \right)^{1/(d-2)},
\quad
P_c =\frac{A(d-2)}{d}\left( {\frac{2A}{d(d-1)B}} \right)^{2/(d-2)}.
\end{equation}
Combining (\ref{eq29}) and (\ref{eq27}) we can get
\begin{equation}
\label{eq30}
\chi x^{d-1}f^{(d-1)/(d-2)}(x)\frac{2A(d-2)}{(d-1)}\left(
{\frac{2A}{d(d-1)B}}
\right)^{1/(d-2)}=Af(x)x^{d-2}(1+x)-B\frac{1-x^d}{1-x}.
\end{equation}
For a fixed $\chi $, i.e. a fixed $T_0 $, we can get a certain
$x$ from Eq. (3.10), and then according to Eqs. (\ref{eq25}) and (\ref{eq26}), the $v_2 $ and $P_0
$ are solved. The corresponding $v_1 $ can be got from $x=v_1 /v_2 $.
Join the points $(v_1 ,P_0 )$ and $(v_2 ,P_0 )$ on isotherms in $P-v$
diagram, which generate an isobar representing the process of isothermal
phase transition or the two phase coexistence situation like that of van der
Waals system. Fig.2 shows the isobars on the background of isotherms at
different temperature and the boundary of the two-phase equilibrium
region by the dot-dashed curve as $n=5$, $b=1$, $Q=1$, $\alpha =0.01$.
The isothermal phase transition process becomes shorter as the temperature
goes up until it turns into a single point at a certain temperature, which
is critical temperature, and the point corresponds to critical state of the
charged topological dilaton AdS black hole.

\begin{figure}
  \centering
  \includegraphics[width=4in]{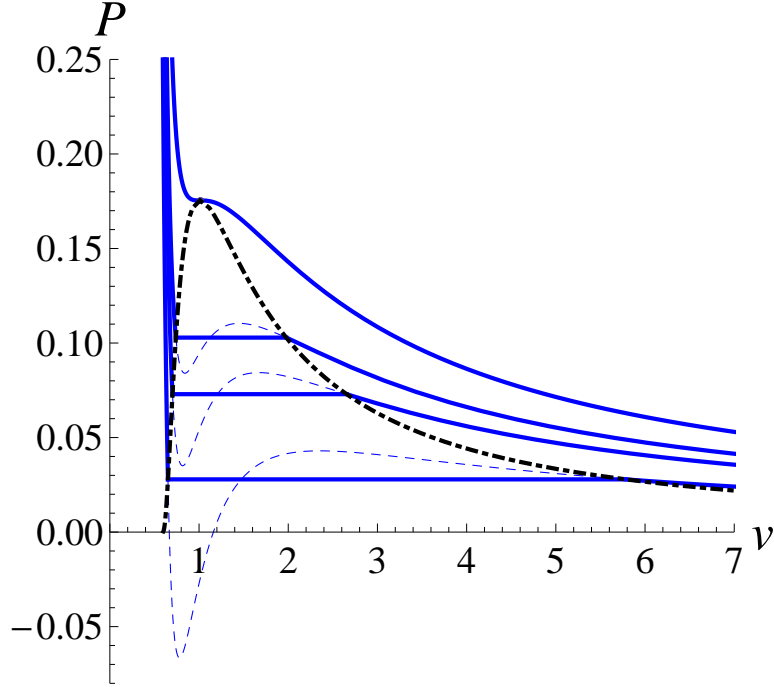}\\
  \caption{\it The simulated isothermal phase transition by isobars and the boundary
of two phase coexistence region for the topological dilaton black hole as
$n=5$, $b=1$, $Q=1$, $\alpha =0.01$.}\label{coexist}
\end{figure}

To analyze the effect of parameters $\alpha $ and $b$ on the phase
transition processes, we take $\chi =0.1,\;0.3,\;0.5,\;0.7,\;0.9$, and
calculate the quantities $x$, $v_2 $, $P_0 $ as $\alpha =0.1,\;0.3,\;0.5$ and
$b=0.2,\;20,\;50$ respectively when $d=5$, $Q=1$. The results are shown in
Table 1.

\begin{table}[htbp]
\caption{\it State quantities at phase transition endpoints with different parameters $\alpha$ and $b$ as $d=5$, $Q=1$}
\begin{center}
\begin{tabular}{|p{20pt}|p{20pt}|p{40pt}|p{40pt}|p{40pt}|p{40pt}|p{40pt}|p{40pt}|p{40pt}|p{40pt}|p{40pt}|}
\hline
& &
\multicolumn{3}{|p{120pt}|}{$\alpha =0.1$} &
\multicolumn{3}{|p{120pt}|}{$\alpha =0.3$} &
\multicolumn{3}{|p{120pt}|}{$\alpha =0.5$}  \\
\hline
$b$&
$\chi $&
$x$&
$v_2 $&
$P_0 $&
$x$&
$v_2 $&
$P_0 $&
$x$&
$v_2 $&
$P_0 $ \\
\hline
\raisebox{-6.00ex}[0cm][0cm]{$0.2$}&
0.9&
0.531&
1.49&
0.145&
0.546&
1.36&
0.220&
0.577&
1.13&
0.502 \\
\cline{2-11}
 &
0.7&
0.266&
2.62&
0.0770&
0.279&
2.36&
0.118&
0.308&
1.92&
0.274 \\
\cline{2-11}
 &
0.5&
0.114&
5.70&
0.0295&
0.121&
5.06&
0.0458&
0.139&
4.00&
0.109 \\
\cline{2-11}
 &
0.3&
0.0253&
24.2&
0.00481&
0.0279&
20.9&
0.00771&
0.0340&
15.6&
0.0195 \\
\cline{2-11}
 &
0.1&
6.32E-5&
9.28E3&
4.55E-6&
8.25E-5&
6.78E3&
8.64E-6&
1.40E-4&
3.65E3.&
3.07E-5 \\
\hline
\raisebox{-6.00ex}[0cm][0cm]{$20$}&
0.9&
0.531&
1.58&
0.128&
0.546&
2.32&
0.0753&
0.577&
4.67&
0.0295 \\
\cline{2-11}
 &
0.7&
0.266&
2.79&
0.068&
0.279&
4.03&
0.0404&
0.308&
7.91&
0.0161 \\
\cline{2-11}
 &
0.5&
0.114&
6.06&
0.0261&
0.121&
8.65&
0.0157&
0.139&
16.5&
0.00640 \\
\cline{2-11}
 &
0.3&
0.0253&
25.7&
0.00426&
0.0279&
35.7&
0.00264&
0.0340&
64.3&
0.00115 \\
\cline{2-11}
 &
0.1&
6.32E-5&
9.87E3&
4.02E-6&
8.25E-5&
1.16E4&
2.95E-6&
1.40E-4&
1.50E4&
1.80E-6 \\
\hline
\raisebox{-6.00ex}[0cm][0cm]{$50$}&
0.9&
0.531&
1.60&
0.125&
0.546&
2.58&
0.0608&
0.577&
6.19&
0.0168 \\
\cline{2-11}
 &
0.7&
0.266&
2.82&
0.0665&
0.279&
4.49&
0.0326&
0.308&
10.5&
0.00915 \\
\cline{2-11}
 &
0.5&
0.114&
6.13&
0.0254&
0.121&
9.63&
0.0126&
0.139&
21.9&
0.00364 \\
\cline{2-11}
 &
0.3&
0.0253&
26.1&
0.00415&
0.0279&
39.7&
0.00213&
0.0340&
85.3&
6.52E-4 \\
\cline{2-11}
 &
0.1&
6.32E-5&
9.99E3&
3.93E-6&
8.25E-5&
1.29E4&
2.39E-6&
1.40E-4&
1.99E4&
1.03E-6 \\
\hline
\end{tabular}
\label{tab1}
\end{center}
\end{table}

From Table 1, we can see that $x$ is unrelated to $b$ but it is
incremental with $\chi$ and $\alpha$. $v_2 $ increases with increasing $b$,
but decreases with increasing $\chi$ and $\alpha $ . $P_0 $ is incremental with $\chi
$ and $\alpha $, but decreases with increasing $b$ . So phase transition process become shorter
with increasing $\alpha $, and it lengthens as $b$ increases.

\section{Two-phase coexistent curves and the phase change latent}

Due to lack of knowledge of chemical potential, the $P-T$ curves of two-phase
equilibrium coexistence for general thermodynamic system are usually
obtained by experiment. However the slope of the curves can be calculated
by Clapeyron equation in theory,
\begin{equation}
\label{eq31}
\frac{dP}{dT}=\frac{L}{T(v^\beta -v^\alpha )},
\end{equation}
where the latent heat of phase change  $L=T(s^\beta -s^\alpha )$, $v^\alpha $,
$s^\alpha $ and $v^\beta $, $s^\beta $ are the molar volumes and molar
entropy of phase $\alpha $ and phase $\beta $ respectively. So Clapeyron equation
provides a direct experimental verification for some phase transition theories.

Here we investigate the two phase equilibrium coexistence $P-T$ curves and
the slope of them for the topological dilaton AdS black hole. Rewrite Eqs. (\ref{eq24})
and (\ref{eq25}) as
\begin{equation}
\label{eq32}
P=y_1 (x),
\quad
T=y_2 (x)
\end{equation}
where
\[
y_1 (x)=\left[ {Ax^{d-2}f(x)-B\frac{1-x^{d-1}}{1-x}} \right]/\left[
{x^{d-1}f^{d/(d-2)}(x)} \right]
\]
\begin{equation}
\label{eq33}
y_2 (x)=\left[ {Af(x)x^{d-2}(1+x)-B\frac{1-x^d}{1-x}} \right]/\left[
{x^{d-1}f^{(d-1)/(d-2)}(x)} \right],
\end{equation}

\begin{figure}
  \centering
  \includegraphics[width=6in]{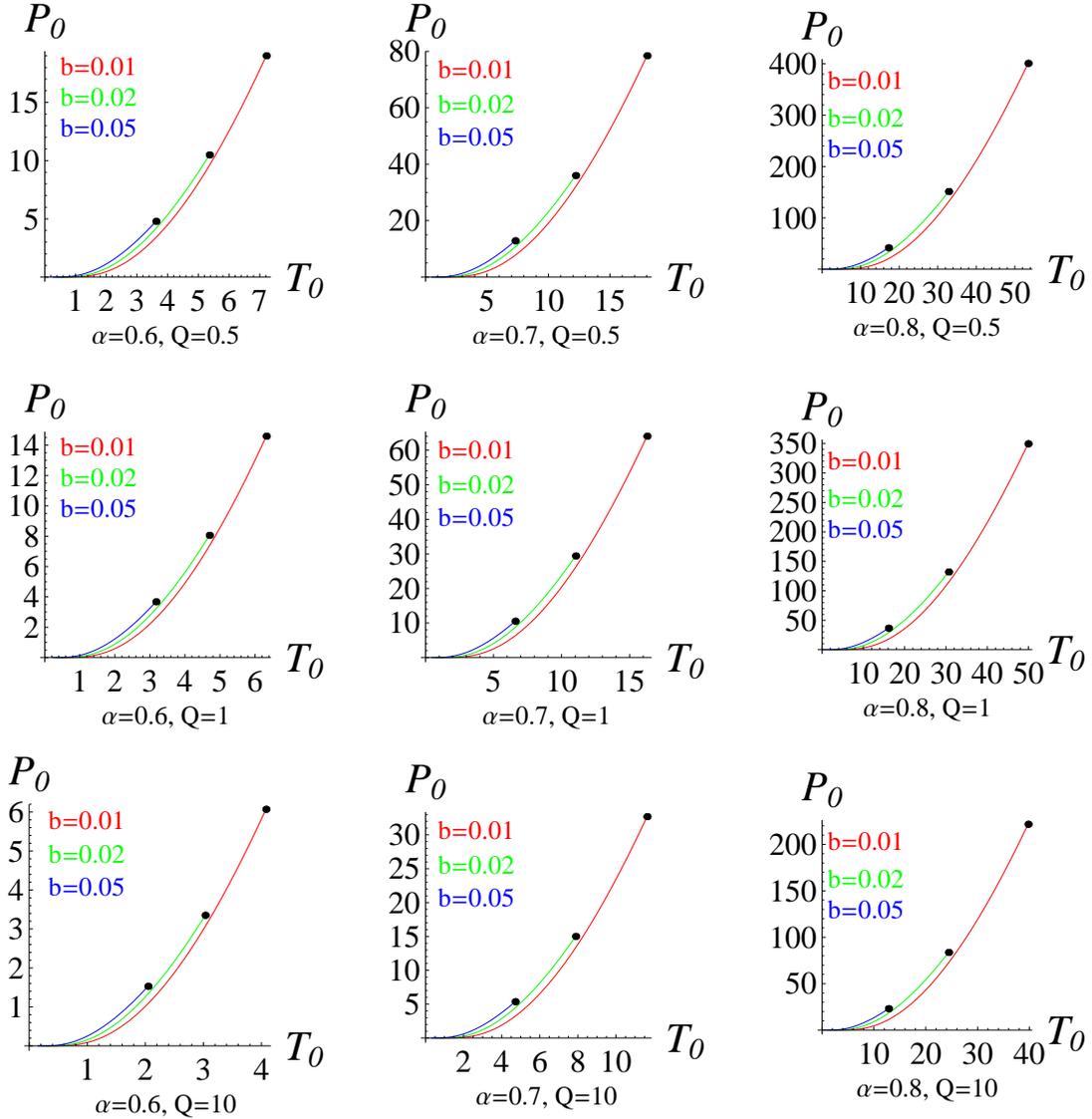}\\
  \caption{\it Two phase equilibrium coexistence curves in $P-T$ diagrams for the
topological dilaton black hole in $5$-dimensional AdS spacetime. In
each diagram, the longest curves (red) correspond to $b=0.01$, the
curves with medium length (green) meet $b=0.02$, and the shortest ones (blue) are with $b=0.05$.}\label{P0-T0}
\end{figure}

we plot the $P-T$ curves with $0<x\le 1$ in Fig.3 when the parameters
$b$, $\alpha $, $Q$ take different values respectively. The curves
represent two-phase equilibrium condition for the
topological dilaton AdS black hole and the terminal points of the
curves represent corresponding critical points.

Fig.3 shows that for fixed $\alpha$ and $Q$, both
the critical temperature and critical pressure decrease as $b$ increases.
Both critical pressure and temperature are
incremental with $\alpha$, but two-phase equilibrium pressure decreases with increasing $\alpha$ at certain
temperature. The change of two-phase equilibrium curve
with parameter $Q$ is similar to that with parameter $b$. As $Q$
becomes larger the critical pressure and critical temperature become smaller, but
at certain temperature the corresponding pressure on $P-T$ curves is
larger for larger $Q$.

From Eq.(\ref{eq33}), we obtain
\begin{equation}
\label{eq34}
\frac{dP}{dT}=\frac{y_1 '(x)}{y_2 '(x)},
\end{equation}
where $y'(x)=\frac{dy}{dx}$. The Eq. (\ref{eq34}) represents the
slope of two-phase equilibrium $P-T$ curve as function of $x$.

From Eqs.(\ref{eq31}) and (\ref{eq34}) we can get the latent heat of phase change
as function of $x$ for $n+1$-dimensional charged topological dilaton AdS black hole,
\begin{equation}
\label{eq35}
L=T(1-x)\frac{y_1 '(x)}{y_2 '(x)}f^{1/(d-2)}(x)=(1-x)\frac{y_1 '(x)}{y_2
'(x)}y_2 (x)f^{1/(d-2)}(x).
\end{equation}
The rate of change of latent heat of phase change with temperature for some usual
thermodynamic systems
\begin{equation}
\label{eq36}
\frac{dL}{dT}=C_P^\beta -C_P^\alpha +\frac{L}{T}-\left[ {\left(
{\frac{\partial v^\beta }{\partial T}} \right)_P -\left( {\frac{\partial
v^\alpha }{\partial T}} \right)_P } \right]\frac{L}{v^\beta -v^\alpha },
\end{equation}
where $C_P^\beta$ and $C_P^\alpha$ are molar heat capacity of phase $\beta$
and phase $\alpha $. For $n+1$-dimensional charged topological dilaton AdS
black hole, the rate of change of latent heat of phase transition with temperature
can be obtained from Eqs.(\ref{eq35}) and (\ref{eq32}),
\begin{equation}
\label{eq37}
\frac{dL}{dT}=\frac{dL}{dx}\frac{dx}{dT}=\frac{dL}{dx}\frac{1}{y_2 '(x)}.
\end{equation}

\begin{figure}
  \centering
  \includegraphics[width=6in]{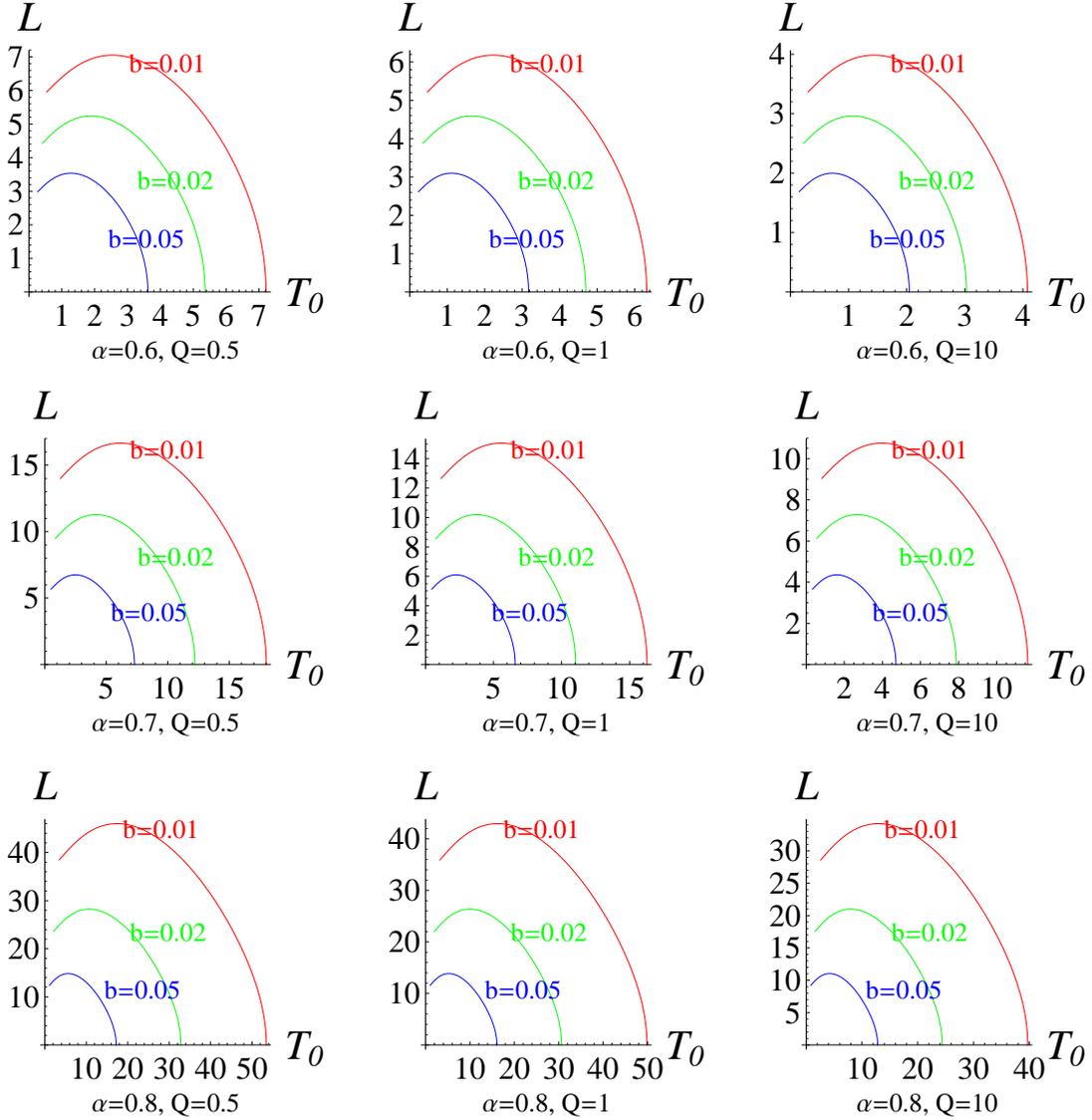}\\
  \caption{\it $L-T$ curves for the topological dilaton black hole in $n$-dimensional
AdS spacetime as $n=5$. In each diagram, the highest curves (red)
correspond to $b=0.01$, the middle curves (green) meet $b=0.02$,
and the lowest curves (blue) are with $b=0.05$.}\label{L-T0}
\end{figure}

Using Eqs. (\ref{eq35}) and (\ref{eq32}) we plot $L-T$ curves in Fig.4 as
the parameters $b$, $\alpha $ and $Q$ take some certain values. From
Fig.4 we can see that the effects of $T$ and the parameters $\alpha
$, $b$, and $Q$ on phase change latent heat $L$. When $T$
increases, $L$ is not monotonous but increases firstly and then
decreases to zero as $T\to T_c $. $L$ decreases with increasing $b$
as other parameters $\alpha$ and $Q$ are fixed. Similarly $L$
decreases with increasing $Q$ for fixed $b$ and $\alpha $. But $L$ is
increment with $\alpha $ for certain $b$ and $Q$. Among the
parameters $b$, $\alpha$ and $Q$, $L$ receives the most effect from
$b$, then $\alpha$, and lastly $Q$.

\section{Discussions and conclusions}

The charged topological dilaton AdS black hole is regarded as a
thermodynamic system, and its state equation has been derived.
But when temperature is below critical temperature,
thermodynamic unstable situation appears on isotherms, and when
temperature reduces to a certain value the negative pressure
emerges, which can be seen in Fig.1 and Fig.2. However, by Maxwell
equal law we established an phase transition process and the problems can be
resolved. The phase transition process at a defined temperature
happens at a constant pressure, where the system specific
volume changes along with the ratio of the two coexistent phases.
According to Ehrenfest scheme the phase transition belongs to the
first order one. We draw the isothermal
phase transition process and depict the boundary of two-phase
coexistence region in Fig.2.

Taking black hole as an thermodynamic systems, many investigations show
the phase transition of some black holes in AdS spacetime and dS
spacetime is similar to that of van der Waals-Maxwell liquid-gas
system\cite{Guna,David,Zhao,Zhao1,Zhao2,Ma,Zhang,Ma1,Ma2,Heidi,Cai,Zou,Zou1,Wei1,Poshteh,Xu,Liu,Xu1},
and the phase transition of some other AdS black hole is alike to
that of multicomponent superfluid or superconducting
system\cite{Frass,Altami,Altami1,Altami2}. It would make sense 
if we can seek some observable system, such as van der Waals
gas, to back analyze physical nature of black holes by their similar
thermodynamic properties. That would help to further understand the
thermodynamic quantities, such as entropy, temperature, heat capacity and
so on, of black hole and that is significant for improving
self-consistent thermodynamics theory of black holes.

The Clapeyron equation of usual thermodynamic system agrees well with
experiment result. In this paper we have plotted the two-phase equilibrium
curves in $P-T$ diagrams, derived the slope of the curves, and
acquired information on latent heat of phase change by Clapeyron equation,
which could create condition for finding some usual thermodynamic systems
similar to black holes in thermodynamic properties and provide theoretical
basis for experimental research on analogous black holes.

\begin{acknowledgments}\vskip -4mm
This work is supported by NSFC under Grant
Nos.(11475108,11175109;11075098), by the Shanxi Datong University
doctoral Sustentation Fund Nos. 2011-B-03, China, Program for the
Innovative Talents of Higher Learning Institutions of Shanxi, and
the Natural Science Foundation for Young Scientists of Shanxi
Province,China (Grant No.2012021003-4).

\end{acknowledgments}

\end{document}